\documentclass[10pt,aps,pra,twocolumn,showpacs,superscriptaddress]{revtex4-2}
\usepackage[english]{babel} 	
\usepackage[usenames, dvipsnames]{color} 
\usepackage{graphicx} 
\usepackage{bm} 
\usepackage{amsmath} 
\usepackage{amsthm} 
\usepackage{amssymb} 
\usepackage{times} 
\usepackage[pdftex,colorlinks=true,urlcolor= blue,linkcolor=Blue,citecolor=RedViolet]{hyperref}
\usepackage{multirow}
\usepackage{graphicx}
\usepackage{bm}
\usepackage{physics}
\usepackage{bbold}
\usepackage{comment}

\providecommand{\bra}[1]{\langle #1 |}
\providecommand{\ket}[1]{| #1 \rangle}

\providecommand{\ketbra}[2]{|  #1\rangle \langle #2 |}

\usepackage{soul}

\begin{document}

\title{Thermodynamics of the optical pumping process in Nitrogen-Vacancy centers}

\author{Ivan Medina}
\email{ivan.medina@ifsc.usp.br}
\affiliation{Instituto de F\'{i}sica de S\~{a}o Carlos, Universidade de São Paulo, CP 369, 13560-970 São Carlos, Brazil}
\affiliation{School of Physics, Trinity College Dublin, Dublin 2, Ireland}
\author{Sérgio R. Muniz}
\affiliation{Instituto de F\'{i}sica de S\~{a}o Carlos, Universidade de São Paulo, CP 369, 13560-970 São Carlos, Brazil}
\author{Elisa I. Goettems}
\affiliation{Instituto de F\'{i}sica de S\~{a}o Carlos, Universidade de São Paulo, CP 369, 13560-970 São Carlos, Brazil}
\affiliation{Department of Chemistry, Bar-Ilan University, Israel}
\author{Diogo O. Soares-Pinto}
\affiliation{Instituto de F\'{i}sica de S\~{a}o Carlos, Universidade de São Paulo, CP 369, 13560-970 São Carlos, Brazil}

\begin{abstract}
The optical pumping process is a fundamental tool for quantum protocols using the electronic and nuclear spins in the Nitrogen-Vacancy (NV) centers platform. In this work, we explore the process of electronic spin polarization through optical pumping from a thermodynamic perspective. We identify the sources of work and heat and show that the heat current is direct related to the experimentally accessible fluorescence of the NV center. We also show that the polarization of the electronic spin depends on the amount of work that is provided to the system by the laser pump. Moreover, we study the von Neumann entropy change in the process. We demonstrate that the entropy can be separated into two contributions: one due to the heat produced and the other due to the work provided, which is a consequence of the non-unitary nature of the pumping process. Finally, we show that increasing the power of the laser results in the increasing of the entropy of the final state, which hinders the polarization efficiency.  
\end{abstract}

\maketitle

\section{Introduction}

Nitrogen-Vacancy (NV) centers in diamond~\cite{jelezko2006single,Doherty2013} have emerged as a prominent platform for solid state quantum technologies. Most part of its success comes from the fact that it has a very long coherence time, being able to reach the order of seconds even at room temperature~\cite{Maurer2012,Bradley2019}. Other key aspects are the optical properties of the NV center, which allow for the electronic spin read out and initialization; coherent control is also achieved by using pulses in the Radio-Frequency domain. Due to these properties, NV centers find a multitude of interesting applications. They can be used for quantum information processing~\cite{wrachtrup2006processing,liu2018quantum,childress2013diamond} and are also a suitable platform for building diamond-based integrated quantum photonic architectures~\cite{lenzini2018diamond}.Moreover, NV centers can be used as extremely precise quantum sensors~\cite{rembold2020introduction,mzyk2022relaxometry}. Its robustness to sense physical quantities, such as magnetic fields~\cite{Dumeige2013,barry2020sensitivity,rondin2014magnetometry}, makes it a powerful tool which finds important applications in bio-chemistry contexts \cite{schirhagl2014nitrogen,zhang2021toward}, where it can be used, for instance, for nanorheometry and nanothermometry of living cells inside complex environments \cite{Gu2023}.

In all those applications, the optical properties of the NV center plays a significant role. The electronic spin of the NV center can be optically excited from its ground state by using a green (532 nm) laser. A spin-conserving decay from the excited state is responsible for the NV spin-dependent fluorescence (637 nm - red), which allows for the direct read out of the electronic spin state. Due to the existence of an additional decaying route via Inter System Crossing (ISC), the optical dynamics can also be used to polarize the electronic spin~\cite{Robledo2011,Goldman2017}. This optical pumping process is then a fundamental tool to control quantum technologies based on NV centers and have been thoroughly explored in the past few years~\cite{jelezko2001spectroscopy,Manson2006,Robledo2011,Tetiene2012,Goldman2015,chakraborty2017polarizing,Goldman2017,gali2019ab,Magaletti2024,Song2020}. 
\begin{figure}[t]
	\centering
	\includegraphics[width=0.48\textwidth]{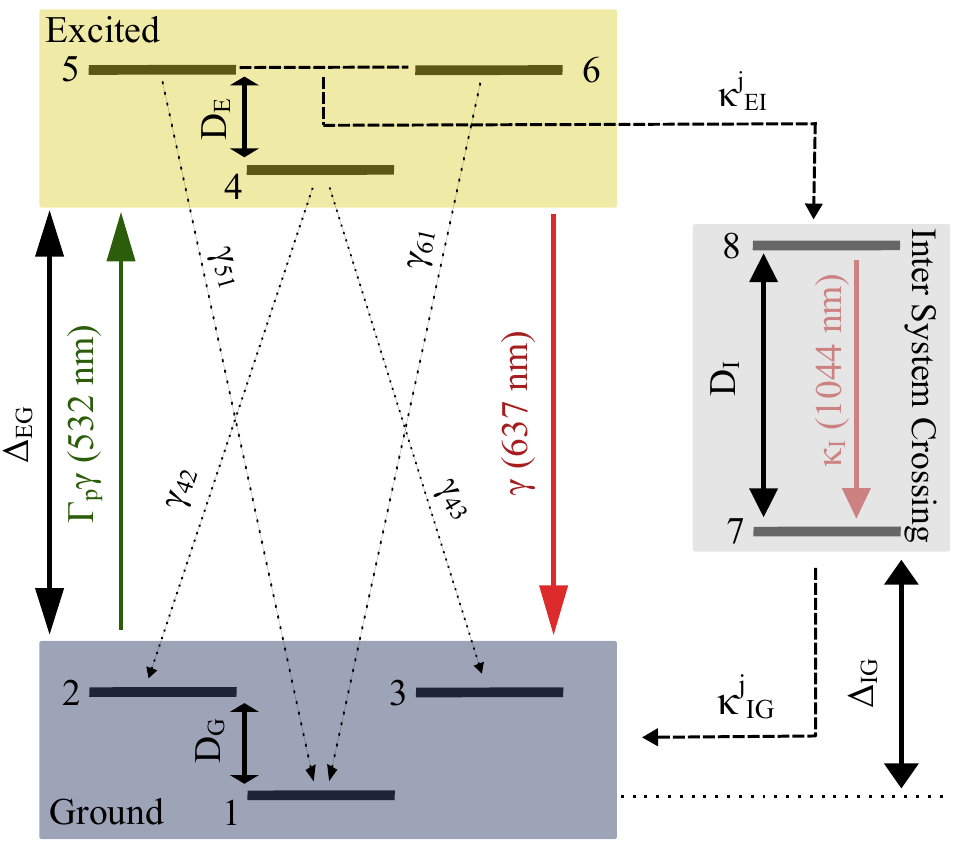}
	\caption{Eight-level model for describing the optical dynamics of the NV center. The model is divided into three subspaces, the ground spin triplet (G), the excited spin triplet (E), and the Inter System Crossing (ISC or I) spin singlet.}
	\label{optpump}
\end{figure}

All these studies focused on understanding the electronic structure and main dissipative mechanisms during the optical excitation of the NV.  Although it is well understood now and routinely used in the most of the experiments, the spin polarization via optical pumping is an non-trivial phenomenon which involves energy transfer from light to the NV center: the electronic spin-state which is initially in a mixed state absorbs energy from the light and is irreversibly driven to a final state that is almost pure. Such an intriguing phenomenon calls for a thermodynamic analysis, which, to the best of our knowledge, has not yet been conducted. Inspired by this, in this paper we investigate the optical polarization of the NV electronic-spin from the thermodynamic perspective. We explore the energy  dynamics of the system and identify the key thermodynamic quantities, heat and work, providing analytical expressions that describe the system's polarization from a thermodynamic perspective. We also explore the von Neumann entropy change of the system and spot its role in the protocol.





This article is organized as follows. In Sec.~\ref{model}, we present the framework used to model the electronic spin of the NV center during the optical process. Specifically, we describe it as an eight-level system that captures the relevant energy states involved in the optical pumping process. In Sec.~\ref{dynamicsnv}, we delve into the optical dynamics of the NV center, which can be effectively described by a Markovian master equation.
In Sec.~\ref{thermodynamics}, we present our main contribution, which is the thermodynamic analysis of the optical pumping process. We define work and heat and relate it to the spin-polarization efficiency. We also investigate how the von Neumann entropy of the system changes during the optical dynamics. We separate the entropy change into two contributions, one due to the irreversible work done on the system and the other due to heat dissipated. Finally, in Sec.~\ref{conc}, we present our final remarks and discuss possible next steps.   

\section{Eight-level Model}\label{model}
\label{pre}

In the NV center system, the optical pumping is primary used to polarize and read out the electronic spin. In this work, we are going to focus on the polarization, in which the optical pump is used to prepare the electronic spin in its ground state. The dynamics of the optical pumping involves three subspaces. The Ground (G) and Excited (E) subspaces, which are spin triplet and the Inter System Crossing (ISC or I for short) subspace, which is a spin singlet. Then, our system can be effectively modeled by an eight-level Hamiltonian, $H$, whose structure is depicted in Fig.~\ref{optpump}. The eigenstates of $H$ are denoted by the orthonormal column vectors $\{\ket{n}\}$ and form an ordered complete basis for the Hilbert space of dimension 8. 

Following Ref.~\cite{Hinks}, we can decompose the total Hilbert space of the system as a direct sum of the three subspaces 
\begin{align}
\mathcal{H}=\mathcal{H}_{\rm G}\oplus\mathcal{H}_{\rm E}\oplus\mathcal{H}_{\rm I},
\end{align}
where ${\rm dim}(\mathcal{H}_{\rm G})={\rm dim}(\mathcal{H}_{\rm E})=3$ and ${\rm dim}(\mathcal{H}_{\rm I})=2$. Then, the total Hamiltonian of the model can be cast in the block diagonal form as
\begin{align}
	H=H_{\rm G}\oplus\mathbf{0}_{\rm E}\oplus\mathbf{0}_{\rm I} +\mathbf{0}_{\rm G} \oplus H_{\rm E}\oplus\mathbf{0}_{\rm I}+\mathbf{0}_{\rm G}\oplus \mathbf{0}_{\rm E}\oplus H_{\rm I},
\end{align}
where $\mathbf{0}_k$ is the null matrix of ${\rm dim}(\mathcal{H}_{\rm k})$ and
\begin{align}
&H_{\rm G}=D_{\rm G} S_z^2+\gamma B_z S_z,\\
&H_{\rm E}=D_{\rm E} S_z^2+\gamma B_z S_z^2+\Delta_{{\rm EG}}S_0,\\
&H_{\rm I}=\left(\Delta_{\rm IG}+\frac{D_{\rm I}}{2}\right)\sigma_0+\frac{D_{\rm I}}{2}\sigma_z.
\end{align}
The matrices $S_z={\rm diag}(1,0,-1)$ and $\sigma_z={\rm diag}(1,-1)$ are the usual Pauli-z operators for dimensions 3 and 2, respectively. The matrices $S_0={\rm diag}(1,1,1)$ and $\sigma_0 ={\rm diag}(1,1)$ in turn, are the identity matrices. The static magnetic field $B_z$ is used to break the energy degeneracy between the states $\ket{2}$ and $\ket{3}$ and $\gamma=28.0$ GHz T$^{-1}$ is the gyromagnetic ratio of the electronic spin. The energy splitting constant $D_{\rm G}=2.87$ GHz is the zero field ($B_z=0$) energy separation between  $\ket{1}$ and the degenerate level $\ket{2}$ and $\ket{3}$; in the same way $D_{\rm E}=1.40$ GHz is the zero field splitting between $\ket{4}$ and the degenerate levels $\ket{5}$ and $\ket{6}$. The energy gap $\Delta_{\rm EG}=4.7 \times 10^{2}$ THz is the effective energy separation between the subspaces G and E. The energy of the I level, $\ket{7}$, is characterized by the energy separation between G and I, $\Delta_{\rm IG}=1.69 \times 10^2$ THz, and the energy separation between the levels $\ket{7}$ and $\ket{8}$ is $D_{\rm I}=2.88 \times 10^2$ THz. The decay parameters used for the simulations in this work  taken from Ref.~\cite{Hinks}, and are presented in Appendix~\ref{appendixa}.

\section{Optical Dynamics and Electronic Spin Polarization}\label{dynamicsnv}
\begin{figure*}[ht]
	\centering
\includegraphics[width=\textwidth]{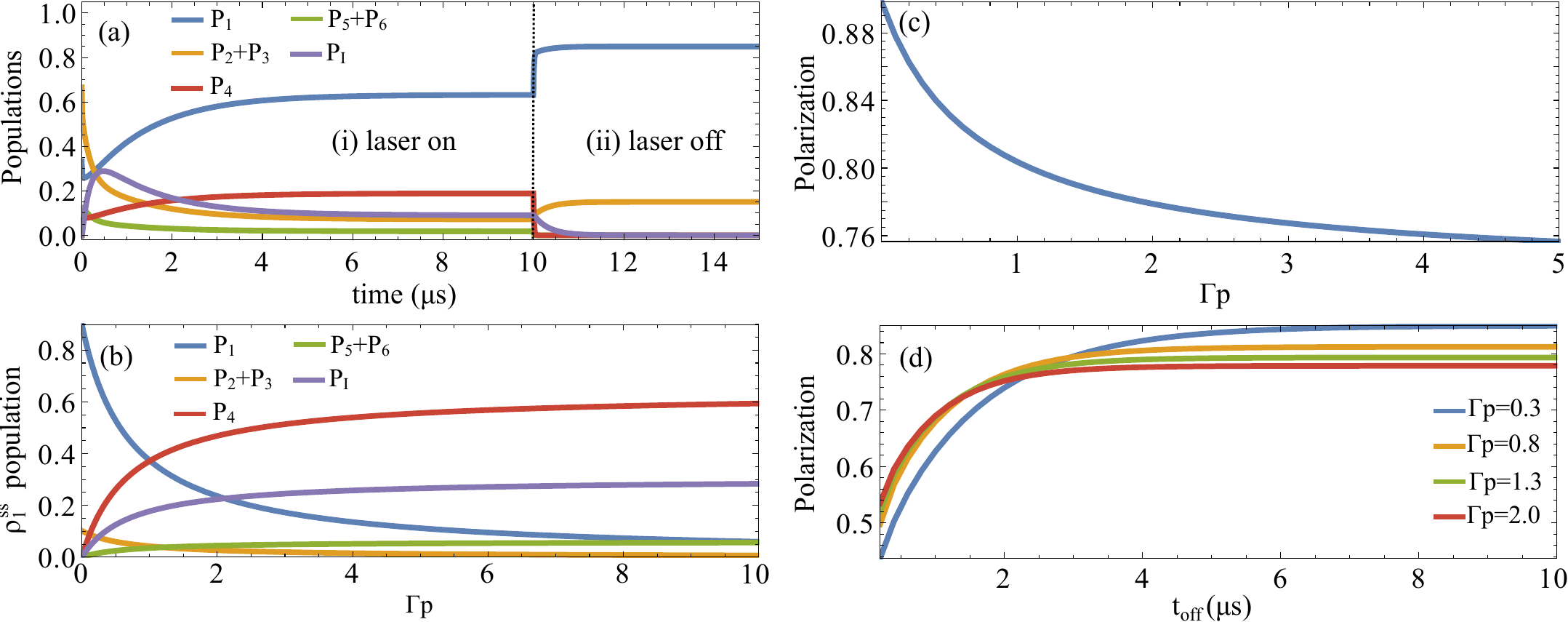}
	\caption{(a) Dynamics of the 8-level system during the optical pumping process in the NV center. The first step consists in turn on the laser and wait until the system reaches the non-equilibrium steady state (NESS) $\rho_1^{\rm ss}$. Then, the laser is turned off and reach a new NESS $\rho_2^{\rm ss}$. The population accumulated in $P_4$ is transferred to $P_1$ polarizing the system. Due to the decaying channel $\kappa_{\rm IG}$ which transfers the population accumulated in the ISC almost equally to the G subspace, the polarization is not perfect. (b) NESS state $\rho_1^{\rm ss}$ in the step ($i$) as a function of the parameter $\Gamma_p$. (c) Polarization as the NESS, $\rho_2^{\rm ss}$, population $P_1$ on step ($ii$) as a function of $\Gamma_p$. (d) Polarization as a function of the time the laser remains on during step ($i$).}
	\label{dynamics}
\end{figure*}
The optical dynamics is triggered when the system is pumped by a green laser pulse ($\lambda = 532$ nm). The absorption of photons takes the system from G to E. The system then relaxes back to G via two main paths. The first is a spin-conserving spontaneous emission. This process is radiative and emits red photons ($\lambda = 637$ nm)) that are responsible for the NV center fluorescence. The second is via an Inter System Crossing (ISC) path, which is responsible for the spin polarization~\cite{Robledo2011,Goldman2017}. In this non-spin conserving route, the population is preferentially transferred from the levels $\ket{5}$ and $\ket{6}$ to $\ket{8}$. Then, the system rapidly decays from $\ket{8}$ to $\ket{7}$ emitting a photon in the infrared ($\lambda \approx 1042$ nm). Finally, the energy is transferred incoherently from $\ket{7}$ to G, being distributed approximately even over the levels $\ket{1}$,$\ket{2}$, and $\ket{3}$. Other non-radiative decays from the E to G exist, but are negligible, in principle, if compared to the other decay mechanisms. However, since they can be characterized we are not dismissing them for the sake of completeness. The pumping process and all the decay channels are illustrated in Fig.~\ref{dynamics}. 

The optical dynamics of the eight-level model can be effectively described by a Lindblad master equation \cite{Petruccione,Carmichael}
\begin{align}
	\dot{\rho}=-i[H,\rho]+\mathcal{D}_{\rm p}(\rho)+\sum_{i=1}^3\mathcal{D}_{\rm d}^{i}(\rho),\label{meq}
\end{align} 
where $\rho$ is the state of the system and $\dot{\rho} = d\rho/dt$ represents the time derivative. The superoperators $\mathcal{D}(\bullet)$, usually called dissipators, describe the non-unitary part of the dynamics and are defined by
\begin{align}
\mathcal{D}(\bullet)=\sum_j L_j \bullet L_j^\dagger-\frac{1}{2}(L_j^\dagger L_j \bullet+\bullet L_j^\dagger L_j),\label{diss}
\end{align}
with $L_j$ being a jump (or Lindblad) operator, which depends on the physical process considered. We conveniently separate the non-unitary dynamics in two contributions, where $\mathcal{D}_{\rm p}(\rho)$ is the laser pump contribution and $\mathcal{D}_{\rm d}^{i}(\rho)$ are the decaying processes. As we will see in the next section, $\mathcal{D}_{\rm p}(\rho)$ and $\mathcal{D}_{\rm d}^{i}(\rho)$ can be respectively connected to the work and heat production in the optical pumping process. The explicit form of each dissipator and jump operator is presented in Appendix~\ref{appendixa}. In this work, we consider the initial state to be a mixed state $\rho(t=0)=\frac{1}{3}\sum_{n=0}^3\ketbra{n}{n}$. As the states in the E and ISC subspaces have energies up to hundreds of THz, they are not thermally excited. For the G subspace the energy levels are of the order of a few GHz. This means that except for very low temperatures, this choice of $\rho(0)$ represents a thermal equilibrium state for the NV center. Since the initial state is diagonal (incoherent) in the basis $\{\ket{n}\}$ and there are no coherence generate terms in the Hamiltonian, 
the system's dynamics will be completely incoherent, i.e., only the populations, $P_n=\bra{n}\rho\ket{n}$, of $\rho$ will evolve in time.

The NV electronic spin polarization is performed in two steps. In the step ($i$), the laser is turned on and interacts with the system until it reaches a first steady state, namely $\rho_1^{\rm ss}$. We remark here that $\rho_1^{\rm ss}$ is a non-equilibrium steady state (NESS). In the step ($ii$), after reaching the first NESS $\rho_1^{\rm ss}$, the laser is turned off, and the system relaxes to a new NESS $\rho_2^{\rm ss}$. The population accumulated in the excited state $\ket{4}$ is almost entirely transferred to $\ket{1}$, finishing the polarization process. For our purposes, by polarization we mean the population of the ground state, $\ket{1}$, in the second NESS $\rho_2^{\rm ss}$. In Fig.~\ref{dynamics} (a), we give an example of the NV optical dynamics. In the plot, we can see the system reaching the NESS $\rho_1^{\rm ss}$ around $t=10$$\mu$s. Thus, the laser is turned off and the population $P_1$ increases to near unity in the NESS $\rho_2^{\rm ss}$. As observed, the polarization is not perfect, with a portion of the population transferring to $P_2$ and $P_3$. This is primarily due to a portion of the population accumulated in the ISC subspace, represented by $P_{\rm I}=P_{\rm 7}+P_{\rm 8}$, that is approximately equally distributed to the ground state triplet when the laser is turned off. As the population decaying rate from $\ket{8}$ to $\ket{7}$ is extremely fast (approximately 1 GHz), we have that $P_{\rm I}\approx P_7$ in the NESS $\rho_1^{\rm ss}$. 

In our model, the power of the laser is related to the unitless parameter $\Gamma_{\rm p}$ (see Appendix~\ref{appendixa}), which is essentially connected to the number of photons that are absorbed by the system. Without loss of generality, from this point forward, we will refer to $\Gamma_{\rm p}$ as laser power. A question that arises is how the polarization depends on the laser power $\Gamma_{\rm p}$. It is natural to expect that if we have a higher $\Gamma_{\rm p}$, this will speed up the dynamics since the pumping excitation rate will increase. Thus, this enables the system to reach the NESS $\rho_1^{\rm ss}$ more quickly. However, it is not immediately clear whether this will enhance the polarization. To gain deeper insight into how the polarization depends on the laser power, we first examine the dependence of the NESS $\rho^{\rm ss}_1$ on $\Gamma_{\rm p}$.
Fig.~\ref{dynamics} (b) illustrates the behavior of the NESS populations $\rho_1^{\rm ss}$ as a function of $\Gamma_{\rm p}$. The plot clearly shows that the NESS population $P_1$ decreases with $\Gamma_{\rm p}$, while the populations $P_4$ and $P_{\rm I}=P_7+P_8$ increase. This occurs because, as the pumping rate $\Gamma_{\rm p}$ increases, the excited subspace populates faster than it decays to the G triplet. In particular, for $\Gamma_{\rm p}>1$ the rate at which the system is excited from E to G is higher than the rate of the spin-conserving decay. This means that for this regime, the subspace E is more populated than the G subspace. We also observe that the system reaches a saturation regime with $\Gamma_{\rm p}$, where the populations in the NESS $\rho^{\rm ss}_1$ stop changing when increasing the laser power. Moreover, increasing $\Gamma_{\rm p}$ also allows for more population to be transferred to the ISC subspace. Then, after turning the laser off, from Fig.~\ref{dynamics} (c), we see that the polarization decreases with $\Gamma_p$. This happens because more population accumulates in $P_7$ in step ($i$), and, consequently, more population is transferred to $P_2$ and $P_3$ in step ($ii$). Finally, in Fig.~\ref{dynamics} (d) we show the dependence of the polarization on the time at which the laser is turned off, $t_{\rm off}$. It is clear that the laser needs to remain on for a sufficient time for the system to reach the NESS $\rho^{\rm ss}_1$, otherwise, the spin polarization is hindered. We would like to emphasize that the evolution of the system after the laser is turned off, i.e., for $t>t_{\rm off}$, follows the same dynamics as in Eq.~\eqref{meq}, but without the laser contribution $\mathcal{D}_{\rm p}(\rho)$. 

\section{Thermodynamics of the optical pumping}\label{thermodynamics}
\begin{figure*}[t]
	\centering
	\includegraphics[width=1.0\textwidth]{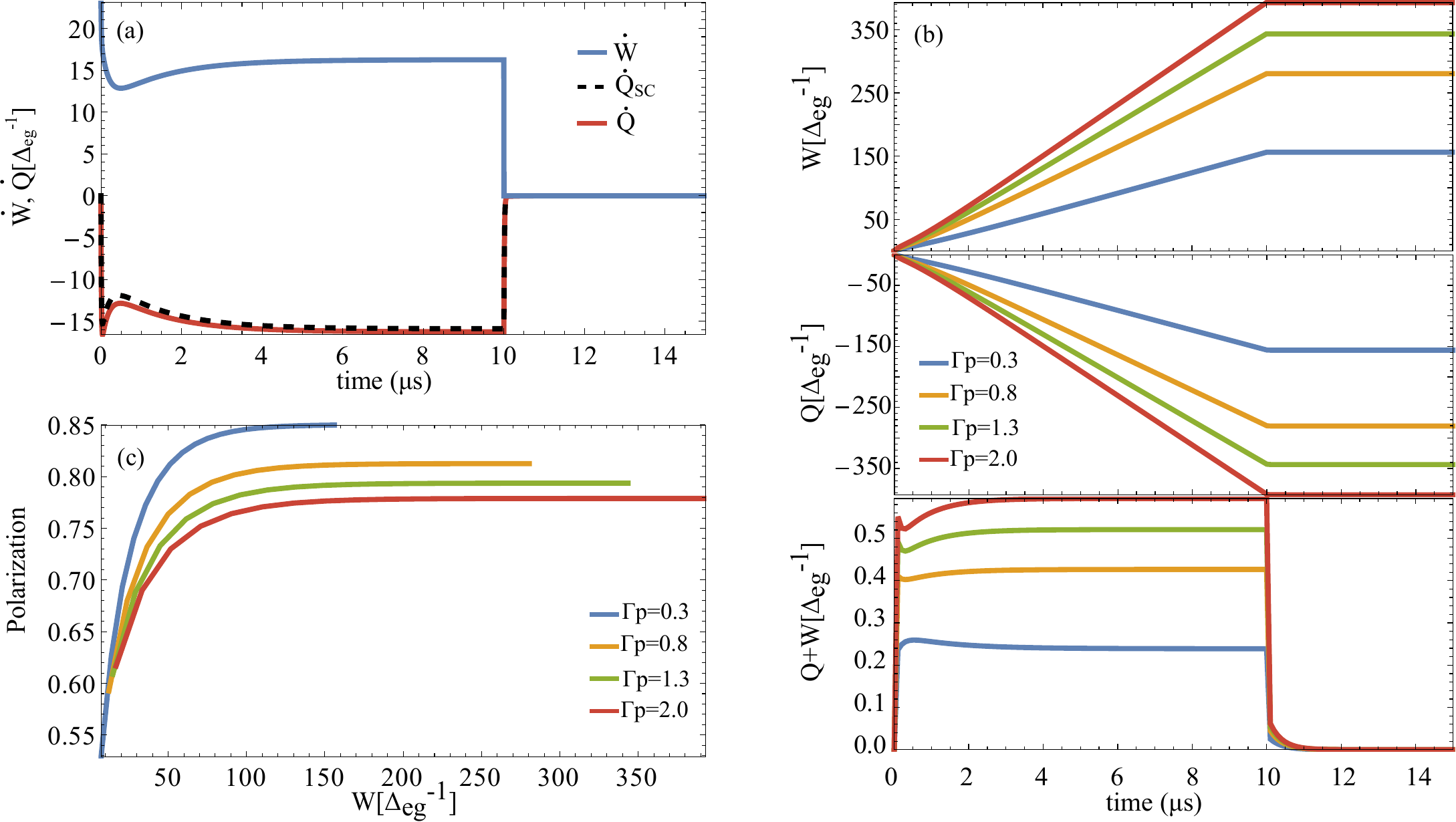}
	\caption{(a) Heat and work fluxes during the optical pumping process. We see that the total heat flux produced $\dot{Q}$ (red-solid line) is almost entirely due to the heat flux produced by the spin-conserving process $\dot{Q}_{\rm sc}$. (b) Heat, work and internal energy variation during the optical pumping process. (c) Parametric plot of the polarization of the NV electronic spin as a function of the work provided by the laser in the step ($i$) of the optical pumping process. To generate this plot we varied the time $t_{\rm off}$ in the interval $(0 ,10]\,\mu s$. }
	\label{thermo}
\end{figure*}

\begin{figure*}[t]
	\centering
	\includegraphics[width=1.0\textwidth]{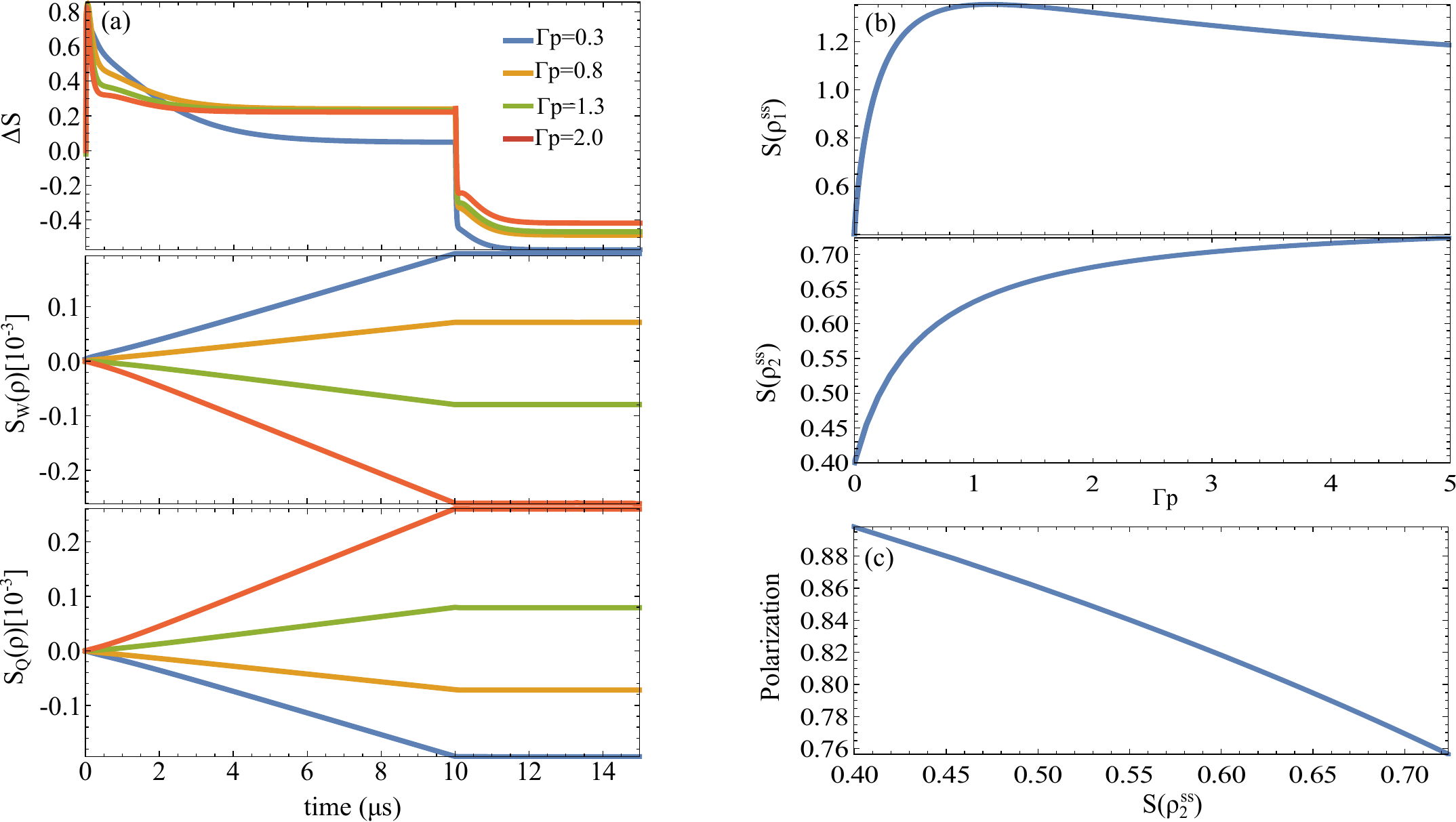}
	\caption{(a) From top to bottom: von Neumann entropy variation $\Delta S=S(\rho(t))-S(\rho(0))$ and entropic contributions $S_W(\rho)$ and $S_Q(\rho)$ as function of time for different values of laser power $\Gamma_{\rm p}$. (b) von Neumann entropy of $\rho_1^{\rm  ss}$ and $\rho_2^{\rm ss}$ as a function of the laser power $\Gamma_{\rm  p}$. (c) Parametric plot of the spin polarization as a function of the von Neumann entropy $S(\rho_2^{\rm 
 ss})$. This plot was generated by varying the laser power in the interval $\Gamma_{\rm p} \in [0,5]$.}
	\label{entropy}
\end{figure*}
In the previous section, we described how the optical dynamics works and how it is used to polarize the spin. This interesting process uses the laser to drive the system from an equilibrium state to a NESS. Then, a specific dissipation process drives the system to a second NESS which is nearly pure. 
Although this process is routinely used in almost all protocols that involve NV centers, to the best of our knowledge, it has not yet been studied from a thermodynamic perspective. From this point of view, the laser acts as a work source that pumps energy incoherently into the system. This energy pump is followed by radiative (photon emission) and non-radiative decaying processes, where energy leaves the system in the form of heat. Interestingly, the energy dissipation occurs in such a way that the final state has less entropy than the initial maximally mixed state. Likewise, the results of the last section suggest that the polarization process depends on amount of work provided to the system through laser pumping [see discussion of Fig.~\ref{dynamics} (d)], which gives us additional motivation for a thermodynamical analysis.

To begin our thermodynamic investigation of the optical pumping process, we first need to define work and heat. As the dynamics of the system is completely incoherent, we can readily identify the work and heat contributions by looking at the time derivative of the system's internal energy, $U(\rho)={\rm tr}\{H\rho\}$, which is given by $\dot{U}(\rho)={\rm Tr}\{H \dot{\rho}\}$. By using the master equation in Eq.\eqref{meq}, we obtain the following expression 
\begin{align}
	\dot{U}(\rho)={\rm Tr}\{H\mathcal{D}_{\rm p}(\rho)\}+\sum_{i=1}^3{\rm Tr}\{H\mathcal{D}_{\rm d}^i(\rho)\}. \label{energyvar}
\end{align}
From Eq.~\eqref{energyvar} and the first law of thermodynamics, we can identify the heat currents $\dot{Q}$ and power $\dot{W}$ as
\begin{align}
	&\dot{W}={\rm Tr}\{H\mathcal{D}_{\rm p}(\rho)\},\label{power1} \\ 
	&\dot{Q}=\sum_i^3\dot{Q}_{i}=\sum_{i=1}^3{\rm Tr}\{H\mathcal{D}_{\rm d}^i(\rho)\}. \label{totalheat}
\end{align}
Working with Eq.~\eqref{power1}, the power delivered by the laser pump is
\begin{align}
	\dot{W}=\Gamma_p \gamma [\Delta_{41}P_{1}+\Delta_{52}P_{2}+\Delta_{63}P_{3}],
\end{align}
where $\Delta_{ij}=\mathcal{E}_i-\mathcal{E}_j$, with $H\ket{n}=\mathcal{E}_n\ket{n}$, is the energy gap between the states $\ket{i}$ and $\ket{j}$. As the energy gap, $\Delta_{\rm EG}$, between the G and E subspaces are on the order of hundreds of THz, and the gap of the energy levels inside the subspaces E and G are a few GHz, we have that $\Delta_{41}\approx\Delta_{52}\approx\Delta_{63}\approx\Delta_{\rm EG}$. We can then rewrite the power as
\begin{align}
	\dot{W}=\Delta_{\rm EG}\Gamma_{\rm p} \gamma P_{\rm G},\label{power}
\end{align}
where $P_{\rm G}$ is the total population of the G subspace, i.e., $P_{\rm G}=\sum_{n=1}^3 P_n$. The work can be obtained by integrating the above quantity in the time interval $[0,t_{\rm off}]$. As observed, the work absorbed by the system is proportional to the G subspace population, with the proportionality constant being the energy gap $\Delta_{\rm EG}$ and the pump excitation rate $ \Gamma_{\rm p} \gamma$. As anticipated, we can control the work done on the system by tuning the controllable parameter $\Gamma_{\rm p}$.

For the heat currents, we can proceed in a similar fashion. Let us focus on the heat contribution coming from the spin-conserving decay part, which we define as $\dot{Q}_{\rm sc}\equiv\dot{Q}_1$. It is straightforward to show that
\begin{align}
	\dot{Q}_{\rm sc}= -\Delta_{\rm EG}\gamma P_{\rm E}, \label{spheat}
\end{align}
where $P_{\rm E}$ is the total population of E, i.e., $P_{\rm E}=\sum_{n=4}^6 P_n$. Equation~\eqref{spheat} implies that the heat current contribution $\dot{Q}_{\rm sc}$ is proportional to the photon current $\gamma P_{\rm E}$, which is exactly the measurable fluorescence of the NV center. In Fig.~\ref{thermo} (a), we show the power and heat current produced in the optical pumping process. We compare the total heat current $\dot{Q}$ obtained from Eq.~\eqref{totalheat} with the heat current obtained from the spin-conserving decay part $\dot{Q}_{\rm sc}$, as in Eq.~\eqref{spheat}. We observe that $\dot{Q}_{\rm sc}$ accounts for almost all the heat current generated in the process. This is a remarkable result as it suggests that we can estimate the total heat current from $\dot{Q}_{\rm sc}$ alone, which is an experimentally accessible quantity. In Fig.~\ref{thermo} (b), we show the heat, work and internal energy, as a function of time for the optical process. We observe that most part of the work and heat are produced in the step ($i$) of the optical process, i.e., while the laser is on, for $t\leq t_{\rm off}=10 \ \mu$s. This is not a surprise for the work contribution, since power is being supplied to the system only while the laser is turned on. For the heat contribution this arises because after the laser is turned off, the system quickly relaxes to the NESS $\rho_2^{\rm ss}$ [see Fig.~\ref{dynamics}]. Since in $\rho_2^{\rm ss}$ the system is entirely in the G subspace, the heat current is zero [see Fig.~\ref{thermo} (a)]. As the heat current variation from $\rho_1^{\rm ss}$ to $\rho_2^{\rm ss}$ occurs very rapidly, the integral of $\dot{Q}$ for $t>t_{{\rm off}}$ is negligible compared to the integral contribution in the interval $t\in[0,t_{\rm off}]$. 

In Fig.~\ref{thermo} (c), we present the polarization as a function of the work delivered by the pump in step ($i$). To generate this plot, we vary the time, ${t_{\rm off}}$, in which the laser is turned off in ($i$) in the interval $t_{\rm off}\in(0\,\mu {\rm s},10\,\mu {\rm s}]$. Then, we compute the polarization and work delivered for each different $t_{\rm off}$. We observe that for a fixed $\Gamma_{\rm p}$, the polarization increases with the work provided until it reaches a saturation regime. The saturation value of the polarization depends on $\Gamma_{\rm p}$. This behavior is consistent with the discussion in Fig.~\ref{dynamics} (d) and with the fact that the longer the laser is on, the more work is pumped into the system. However, it provides a more neat and interesting thermodynamic interpretation. It means that the laser need to be turned on for a sufficient amount of time so the system can absorb enough work to be able to reach the NESS $\rho_1^{\rm ss}$, thus providing the ``maximal polarization". The saturation occurs when the system reaches $\rho_1^{\rm ss}$, for which we have $\dot{W}=\dot{Q}$. This means that in the NESS all the work supplied will leave the system in the form of heat due the decaying processes.

In order to provide a complete thermodynamic perspective of the process, we now study the instantaneous variation of system's von Neumann entropy, which reads $S(\rho)=-{\rm Tr}\{\rho\ln \rho\}$. It is worth remarking that since the dynamics is completely incoherent, in this case the von Neumann entropy is equivalent to the the Shannon entropy for the system's population, i.e., $H(\vec{P})=-\sum_{n=1}^8 P_n\ln P_n$, where $\vec{P}$ is a vector with elements $P_n$. The density operator formalism allows us to define the contributions from  work and heat to the von Neumann entropy change. By taking the time derivative of the von Neumann entropy, a simple calculation yields $\dot{S}(\rho)=- {\rm Tr}\{\dot{\rho}\ln \rho\}$. Then, by using the master equation in Eq.~\eqref{meq}, we can write $\dot{S}(\rho)$ as
\begin{align}
    \dot{S}(\rho)=-{\rm Tr}\{\rho\mathcal{D}_{\rm p}^*(\ln \rho)\}-\sum_{i=1}^{3}{\rm Tr}\{\rho\mathcal{D}_{\rm d}^{i*}(\ln \rho)\},
\end{align}
where $\mathcal{D}^*(\bullet)=\sum_j   L_j^\dagger \bullet L_j-\frac{1}{2}(L_j^\dagger L_j \bullet+\bullet L_j^\dagger L_j)$ is the adjoint of the dissipator $\mathcal{D}(\bullet)$ defined in Eq.~\eqref{diss}. We can now define,
\begin{align}
&\dot{S}_W(\rho)=-{\rm Tr}\{\rho\mathcal{D}_{\rm p}^*(\ln \rho)\}\label{swork},\\
&\dot{S}_Q(\rho)=-\sum_{i}^{3}{\rm Tr}\{\rho\mathcal{D}_{\rm d}^{i*}(\ln \rho)\}.\label{sheat}
\end{align}
In the optical pump protocol, the work is done in an irreversible way (it is not a unitary operation), so it leads to an entropy change in the system that is given by $\dot{S}_{\rm W}(\rho)$, as in Eq.~\eqref{swork}. The term $\dot{S}_Q(\rho)$, in turn, accounts for the entropy change due to heat produced by the decay processes. By integrating $\dot{S}(\rho)$ in time we obtain $\Delta S=S(\rho(t))-S(\rho(0))=S_W(\rho)+S_Q(\rho)$, where $S_W(\rho)$ and $S_Q(\rho)$ are obtained by integrating Eqs.~\eqref{swork} and \eqref{sheat}.

In Fig.~\ref{entropy} (a), we show $\Delta S$ and the contributions $S_Q(\rho)$ and $S_W(\rho)$ as a function of time for different values of $\Gamma_{\rm p}$. We see that in the very the beginning of the dynamics, $\Delta S$ rapidly increases and then it relaxes to a constant value when the system reaches the NESS $\rho^{\rm ss}_1$. After the laser is turned off, $\Delta S$ quickly decays to a new stationary value when the system reaches $\rho^{\rm ss}_2$. We can clearly notice that the value of $\Delta S$ for $\rho^{\rm ss}_1$ and $\rho^{\rm ss}_2$ depends on the laser power $\Gamma_{\rm p}$. It is also interesting to observe that the contributions $S_Q(\rho)$ and $S_W(\rho)$ change signs depending on the value of $\Gamma_{\rm p}$. For $\Gamma_{\rm p} \lessapprox 1.0$, the laser is not strong enough to overpopulate the E subspace. Consequently, the effect of the laser makes the system's sate more mixed, contributing with a positive entropy. In this case the dissipation tries to push the system into the G subspace, thus contributing with a negative entropy. For $\Gamma_{\rm p}\gtrapprox1$, the roles are inverted. As the laser excites the system more rapidly than it can dissipate to G, the population tends to be concentrated more in the E and S subspaces. The dissipation, on the other hand, tends to mix the system by pushing it to the ground state. Consequently, for this regime the energy pumped into the system contributes with a negative entropy, while the energy dissipated as heat contributes with positive entropy. 

In Fig.~\ref{entropy} (b), the von Neumann entropy of the NESSs $\rho_1^{\rm ss}$ and $\rho_2^{\rm ss}$ are shown as a function of $\Gamma_{\rm p}$. We observe that for $\Gamma_{\rm p}\approx1.13$, $S(\rho_1^{\rm ss})$ reaches a maximum. After this maximum point, $S(\rho_1^{\rm ss})$ decreases monotonically . This behavior is explained by the fact that, for $\Gamma_{\rm p}< 1.13$ the population of $\rho_1^{\rm ss}$ is more concentrated in the G subspace and for $\Gamma_{\rm p}> 1.13$ it becomes more concentrated in the E and ISC subspaces. This means that for $\Gamma_{\rm p}\approx1.13$, the populations of the NESS $\rho_1^{\rm ss}$  are as evenly distributed as possible between the three subspaces [see Fig.~\ref{dynamics}]. Finally, despite the non-monotonic behavior of $S(\rho_1^{\rm ss})$ with $\Gamma_{\rm p}$ we see that $S(\rho_2^{\rm ss})$ increases monotonically as we increase the laser power. This explains why the saturation of the polarization with the work provided by the laser is lowered as we increase $\Gamma_{\rm p}$ [see Fig.~\ref{thermo} (c)]. As we increase $\Gamma_{\rm p}$ the entropy of the NESS $\rho^{\rm ss}_2$ also increases, and this hinders the polarization, as we can see from Fig.~\ref{entropy} (c).

\section{Final Remarks}
\label{conc}
In this work, the optical pumping process in NV centers is approached from a thermodynamic perspective. We defined work and heat as the energy currents arising due to the pumping and decaying processes, respectively. This separation of work and heat conforms to the Alicki formalism \cite{Alicki1,Alicki2} when the Hamiltonian is time independent and the work source is an incoherent pumping. It is important to notice that the process described in this work is completely incoherent. From a thermodynamical perspective, we found that the heat produced during the optical pumping process is directly related to the experimentally accessible fluorescence of the NV center. Moreover, our results show that the polarization of the electronic spin depends on the amount of work that is provided by the pump, i.e., to be efficiently polarized, the system needs to absorb enough work from the laser in order to reach a steady state. We also explored how the von Neumann entropy of the system evolves during the optical process. We found that although increasing the laser power makes the system polarizes faster, the entropy of the final state will also be higher, worsening the polarization efficiency. All these interesting result may be connected with the hypothesis of dissipative adaptation, where one of the ingredients is the absorption of work by the system in order for it to  self-assemble or reach a specific non-equilibrium state \cite{England2015,Valente2021,Ganascini2024}. 

It is important to remark that in this work we did not take into account the effects of the environment, that lead to system thermalization. This is justified by the fact that for NV centers at room temperature the characteristic relaxation time $T_1$ is at the order of ms~\cite{Doherty2013}, while the optical process occurs at a few $\mu$s time scale. As we see in Fig.~\eqref{dynamics}, by lowering the power of the laser, one in principle can obtain a better polarization efficiency. However, as the laser power is lowered, the time the system takes to reach the NESS $\rho_1^{\rm ss}$ increases. As consequence, the polarization time also increases. Then, for low values of $\Gamma_{\rm p}$, thermalization effects must be taken into account, since it will play a role as a hindering mechanism on the polarization of the electronic spin. 

Finally, it is worth mentioning that the optical pumping process can also be used to polarize nuclear spins, by using, for example, the Excited State Level Anticrossing (ESLAC) method~\cite{Maze2021}. In this method, high intensity magnetic fields ($B_{z} \approx 500$ G or higher) are used to bring the electronic-spin levels close to the nuclear-spin levels and transfer polarization through the hyperfine coupling. Besides the use of coherent microwave pulses, other mechanisms, such as the ionization of the NV${}^-$ due to effects of charge-state conversion, may play a role on the spin polarization~\cite{Poggiali2017,Wirtitsch2023}. Exploring the thermodynamics of these processes would be an interesting follow up of this work, since the presence of coherence plays a non-trivial role in the definitions of work and heat and is an open question in the field of quantum thermodynamics~\cite{Binder2015,Bernardo2020}.



\begin{acknowledgments}
The authors thank Laetitia Bettmann, Saulo V. Moreira, Frederico Brito, Gabriel T. Landi, Samuel L. Jacob and John Goold  for the fruitful discussions and feedback on our work. I.M. acknowledges financial support from São Paulo Research Foundation - FAPESP (Grants No. 2022/08786-2 and No. 2023/14488-7). S.R.M acknowledges financial support from São Paulo Research Foundation - FAPESP (Grants No. 19/27471-0 and No. 13/07276-1). E.I.G acknowledges Israel Science Foundation (Grant No.1364/21). D.O.S.P acknowledges the support by the Brazilian funding agencies CNPq (Grant No. 304891/2022-3), FAPESP (Grant No. 2017/03727-0) and the Brazilian National Institute of Science and Technology of Quantum Information (INCT/IQ). This study was financed in part by the Coordenação de Aperfeiçoamento de Pessoal de Nível Superior, Brazil (CAPES), Finance Code 001.
\end{acknowledgments}

\bibliographystyle{unsrt}
\bibliography{refs.bib}

\appendix

\section{Jump operators and parameters}\label{appendixa}
In this appendix we explicitly construct all the dissipators, $\mathcal{D}_i(\rho)$, of the Master Equation \eqref{meq}. The first contribution, $\mathcal{D}_{\rm p}(\rho)$, is responsible for the incoherent spin-preserving pumping from the ground state to the excited state. There are three jump operators associated with the pumping
\begin{align}
	&L_1=\sqrt{\Gamma_p \gamma}\ketbra{4}{1},\\
	&L_2=\sqrt{\Gamma_p \gamma}\ketbra{5}{2},\\
	&L_3=\sqrt{\Gamma_p \gamma}\ketbra{6}{3}.
\end{align}
Analogously, the contribution $\mathcal{D}_{\rm sc}\equiv\mathcal{D}_{\rm d}^{1}$, which accounts for the spin-conserving decays, is defined by the jump operators
\begin{align}
	&L_4=\sqrt{ \gamma}\ketbra{1}{4},\\
	&L_5=\sqrt{ \gamma}\ketbra{2}{5},\\
	&L_6=\sqrt{ \gamma}\ketbra{3}{6}.
\end{align}
This process is responsible for the fluorescence of the NV center and is important for the measurement process. The fluorescence  is obtained by multiplying the decaying rate $\gamma$ by the instantaneous population of E, i.e, $\gamma P_{\rm E}$.

The contribution ${D}_{\rm d}^{2}$ is the ISC decay path, which is defined by 7 jump operators
\begin{align}
	&L_7=\sqrt{ \kappa_{\rm EI}^{4}}\ketbra{8}{4},\\
	&L_8=\sqrt{ \kappa_{\rm EI}^{5}}\ketbra{8}{5},\\
	&L_9=\sqrt{ \kappa_{\rm EI}^{6}}\ketbra{8}{6},\\
	&L_{10}=\sqrt{\kappa_{\rm I}}\ketbra{7}{8},\\
	&L_{11}=\sqrt{ \kappa_{\rm IG}^{1}}\ketbra{1}{7},\\
	&L_{12}=\sqrt{ \kappa_{\rm IG}^{2}}\ketbra{2}{7},\\
	&L_{13}=\sqrt{ \kappa_{\rm IG}^{3}}\ketbra{3}{7}.
\end{align}
 The jump operators $L_7$, $L_8$ and $L_9$ are responsible for the decays from E to S; $L_{10}$ is the radioactive decay from $\ket{8}$ to $\ket{7}$, which emits a photon in the infra-red (1042 nm). The last three jump operators describe the decay from S to G. Finally, the last contribution, ${D}_{\rm d}^{3}$, describes small non-spin-conserving decays, and are defined by 4 jump operators
 \begin{align}
 	&L_{14}=\sqrt{ \gamma_{42}}\ketbra{2}{4},\\
 	&L_{15}=\sqrt{ \gamma_{43}}\ketbra{3}{4},\\
 	&L_{16}=\sqrt{ \gamma_{51}}\ketbra{1}{5},\\
 	&L_{17}=\sqrt{ \gamma_{51}}\ketbra{1}{6}.
 \end{align}
All the decay parameters can be experimentally characterized. To show our results, we withdraw the parameters values from Ref.~\cite{Hinks}, which are expressed in Table \ref{par}.
\begin{center}
\begin{table}[h]
	\begin{tabular}{|l|}
		\hline
		{\bf Parameters } \\
		\hline\hline
		\ $\gamma$ = 77.0 MHz \\ 
		\ $\gamma_{42}=\gamma_{43}=\gamma_{51}=\gamma_{61}$ = 0.25 MHz\\ 
		\ $\kappa^{4}_{\rm EI}$ = 0 \\
		\ $\kappa^{5}_{\rm EI}=\kappa^{6}_{ES}$ = 15.0 MHz   \\ 
		\ $\kappa^{1}_{\rm IG}=\kappa^{2}_{\rm IG}=\kappa^{3}_{\rm IG}$ = 1.0 MHz \\ 
		\ $\kappa_{\rm I}$ = 1.0 GHz \\ 
		\hline    
	\end{tabular}
    \caption{Parameters used in this work.}
    \label{par}
    \end{table}
\end{center}
\end{document}